\documentclass{bmcart}

\usepackage[utf8]{inputenc} 
\usepackage{graphicx}
\usepackage[figuresright]{rotating}
\usepackage{array}
\newcolumntype{C}[1]{>{\centering\let\newline\\\arraybackslash\hspace{0pt}}m{#1}}

\startlocaldefs
\endlocaldefs

\begin{document}

\begin{frontmatter}
\begin{fmbox}               
\dochead{Research}


\title{fMRI-based Static and Dynamic Functional Connectivity Analysis for Post-stroke Motor Dysfunction Patient: A Review}


\author[
  addressref={aff1,aff2},                   
  email={kaichaowu@stu.edu.cn}   
]{\inits{KC}\fnm{Kaichao} \snm{Wu}}
\author[
  addressref={aff3},                   
  email={b.jelfs@bham.ac.uk}   
]{\inits{BJ}\fnm{Beth} \snm{Jelfs}}
\author[
  addressref={aff2},                   
  email={katrina.neville@rmit.edu.au}   
]{\inits{KN}\fnm{Katrina} \snm{Neville}}
\author[
  addressref={aff1},
  corref={aff1},
  email={qiangfang@stu.edu.cn}
]{\inits{QF}\fnm{John Q.} \snm{Fang}}

\address[id= aff1]{
  \orgdiv{Department of Biomedical Engineering, College of Engineering},             
  \orgname{Shantou University},          
  \city{Shantou},                              
  \cny{P.R.China}                                    
}
\address[id= aff2]{%
  \orgdiv{School of Engineering},
  \orgname{RMIT University},
  \city{Melbourne},
  \cny{Australia}
}
\address[id=aff3]{%
  \orgdiv{School of Engineering and Physical Sciences},
  \orgname{The University of Birmingham},
  \city{Birmingham},
  \cny{UK}
}



\end{fmbox}  


\begin{abstractbox}

\begin{abstract} 
Functional magnetic resonance imaging (fMRI) has been widely utilized to study the motor deficits and rehabilitation following stroke. In particular, functional connectivity(FC) analyses with fMRI at rest can be employed to reveal the neural connectivity rationale behind this post-stroke motor function impairment and recovery. However, the methods and findings have not been summarized in a review focusing on post-stroke functional connectivity analysis. In this context, we broadly review the static functional connectivity network analysis (SFC) and dynamic functional connectivity network analysis (DFC) for post-stroke motor dysfunction patients, aiming to provide method guides and the latest findings regarding post-stroke motor function recovery. Specifically, a brief overview of the SFC and DFC methods for fMRI analysis is provided, along with the preprocessing and denoising procedures that go into these methods. Following that, the current status of research in functional connectivity networks for post-stoke patients under these two views was synthesized individually. Results show that SFC is the most frequent post-stroke functional connectivity analysis method. The SFC findings demonstrate that the stroke lesion reduces FC between motor areas, and the FC increase positively correlates with functional recovery.  Meanwhile, the current DFC analysis in post-stroke has just been uncovered as the tip of the iceberg of its prospect, and its exceptionally rapidly progressing development can be expected.
\end{abstract}




\end{abstractbox}
%


\end{frontmatter}




\section*{Introduction}

According to the report on Global Burden Disease~\cite{RN3}, stroke is the second leading cause of death and the second largest cause of disability. In a global world, there were 101 million stroke cases and 13.7 million new stroke survivors~\cite{RN2} in 2019, which is nearly five times that of 2013. As more than 60\% of the survivors are left with severe sequela, stroke has become the largest known cause of complex disability~\cite{RN4}, and 77\% of stroke survivors suffer from slow upper limb/hand movement disorder~\cite{RN5}.

There are multiple causes for the quintupling of survivors in such a short period of time, including ageing, growing populations,improved post-stroke care, increased risk factors, to name a few.~\cite{RN2}. With the sharp growth of stroke survivors, effective post-stroke rehabilitation seems to be great necessary. Various restorative therapies can assist stroke survivors in improving their motor function deficits. They mainly include (i) dynamic splints-based therapy (e.g., physiotherapy, restraint-induced kinesiotherapy, gait therapy)~\cite{RN7, RN6}, (ii) electrical muscle stimulation (EMS) therapy~\cite{RN8, RN9}, (iii) device-driven therapy (e.g., robotics, brain-machine interfaces (BCI))~\cite{RN18, RN11, RN13, RN10}, (iv) transcranial magnetic stimulation therapy (TMS)~\cite{RN14, RN15}, and (v) mirror therapy~\cite{RN16, RN17}. The prospects for these therapies seem bright. The hand function, for example, shows signs of recovery even during the chronic recovery phase~\cite{RN13}. However, the effectiveness of these rehabilitation measures varies from individual to individual. Not every patient recovers their limbs to some extent after a stroke;and even for individuals with identical levels of initial functional impairment, the rehabilitation effects vary greatly among people~\cite{RN20, RN19}. Many unknown factors influence the outcome of recovery~\cite{RN22, RN21}. Since the lesions that results in post-stroke disability are located in the brain, to gain insight into these factors, it is crucial to comprehend the process in the brain when function recovery occurs.

Emerging evidence has revealed that post-stroke recovery is a time-depended process. The interrupted connectivity within the central nervous system reorganizes in structure and function over time to adapt to damage caused by a stroke. Hence, a lot of cross-sectional and longitudinal study has been carried out recently to gain insight into this process. By taking advantage of the dynamic alteration of the brain connectivity network, the behaviour of the brain system can recognize, for instance, how recovery adapts structurally and functionally over time and how this adaptation underpins functional recovery~\cite{RN23}. In addition, the structure or function connectivity dynamics in brain network reorganization can help reveal what factors promote or impede the recovery process, thus informing more productive intervening therapies in accordance with this neural network modulation and ultimately facilitating greater rehabilitation programs. 

Structurally, stroke lesions are considered one of the most important factors that cause network disconnection. Interestingly, after stroke, the lesion-surrounding regions can bypass the damage and rebuild connections to support the patient relearn the lost function~\cite{RN21}. For example, one study by~\cite{RN25}  has demonstrated that the corticospinal tract (CST) connecting the cerebellum and the primary motor cortex will be generated after stroke, and the newly found physical connections are associated to skillful motor control. As a pathway connecting the motor cortex with motor neurons projecting from the spinal cord, CST plays an essential role in the motor control system~\cite{RN28}. Once CST is damaged following stroke, the alternative pathways will be recruited to compensate for the lost connections~\cite{RN25} and thus result in further change in structure connectivity~\cite{RN27, RN26}. 
Beyond impairing local physic connectivity, a stroke lesion can also alter the neural interaction between directly or indirectly connected brain areas~\cite{RN30}. This interaction (or communication) is shown with functional collaboration between brain regions, which is also referred to as Functional Connectivity (FC)~\cite{RN37}. Over past few decades, the changes in functional network architecture during stroke recovery have been continually reported. A common finding in numerous studies looking at post-stroke FC is the decreased interhemispheric FC at the initial stage after stroke. Yet, this abnormal FC will develop and return toward the nearly ordinary level~\cite{RN44, RN33, RN34}. And this gradually enhanced FC is demonstrated to favorably correlate with motor recovery in the subacute or chronic phase~\cite{RN36, RN35, RN33}.

Since functional connectivity has become an essential metric in neuroscience related to stroke recovery, a variety of means recording the brain activity have been utilized to explore the brain functional connectivity pattern, including EEG(Electroencephalography), MEG (Magnetoencephalography), and fMRI (Functional Magnetic Resonance Imaging). EEG and MEG measure the FC by recording the electromagnetic neural activity, whilst fMRI achieves this target by measuring the consistency of the blood oxygenation level-dependent (BOLD) signals across brain regions over time.

Among these methods, the fMRI is a well-liked non-invasive imaging method due to its high spatial-temporal resolution~\cite{RN21, RN40}. However, whether the connection is evaluated at rest or when doing a task is a crucial factor that cannot be overlooked in post-stroke FC analysis with fMRI. According to the observations across studies, these two ways often produce variable results. Resting-State (RS)-fMRI has certain advantages compared with task-based fMRI. The first is that studies with RS-fMRI do not need to design the movement paradigm and quantify the motor executive performance, while that is hard part across studies with task-based fMRI and can be greatly diverse. Measure task performance, for example, in hand grip task~\cite{RN42} and hand movement task~\cite{RN43} are different. And importantly, RS-fMRI provides a unified manner for comparison between studies, thus facilitating further understanding of recovery mechanisms. yet, the discovered recovery mechanisms in the study with task-based fMRI may be only effective in particular tasks. Therefore, this review concentrates on research employing rest-state fMRI to examine brain functional connectivity following stroke. In addition, when analyzing the function connectivity from the perspective of the method processing the RS-fMRI data, the distinction between the studies is the time scales over which the brain function connectivity evolves. Conventional methods assume that the FC measures are stationary over a full MRI scan, while it has shown that the FC fluctuates even over the seconds~\cite{RN45}~\cite{RN46} and the static FC network is too simplistic to capture the complete representation of FC evolution~\cite{RN47}. Recently, a growing number of methods have been introduced to explore dynamic functional connectivity following stroke and made an effort to bring an all-new perspective to investigate the recovery mechanism. 

In 2011, the longitudinal changes of resting-state functional connectivity during motor recovery after stroke have been evaluated~\cite{RN37}. The dynamic development of brain connectivity after stroke in 2018 has also been documented~\cite{RN23}. However, a review concentrating on post-stroke functional connectivity analysis from the standpoint of the techniques has not yet provided a summary of the most recent findings. Hence, this review attempts to provide overview of the latest progress in the changes to brain motor functional architecture following stroke from the perspective of static and dynamic functional networks(SFC and DFC). By synthesizing the present understanding of functional networks in the resting state from various viewpoints, this review can provide potential method guides and feasible references to map the post-stroke neuroplasticity of brain circuits. Additionally, the comprehension of resting-state functional network under different analysis methods can provide viable guidance in designing dynamic neural rehabilitation interventions to benefit stroke patients.


\section*{Methodology}
\subsection*{literature search}
The literature search was restricted to English-language articles published between January 2000 and May 2022 in the following electronic databases: PubMed, Web of Science, IEEE Xplore, ScienceDirect, MEDLINE (OvidSP), CDSR (Cochrane database of systematic reviews), Scopus, Compendex, Wiley Online Library, Academic Search Premier, and Springer Link. The electronic search terms were Stroke AND fMRI AND Functional Connectivity AND Motor deficit AND Rehabilitation. Studies that include task-related fMRI and involve effective connectivity and beyond motor function were excluded. Besides, this review included studies that explore the development of functional connectivity analysis methods. Particularly, those studies that employ these approaches for post-stroke motor function impairment and rehabilitation have been covered.

\subsection*{Terms and definition }

\subsubsection*{Post-stroke stage}
\begin{figure}[ht]
    \includegraphics[width =0.95\textwidth]{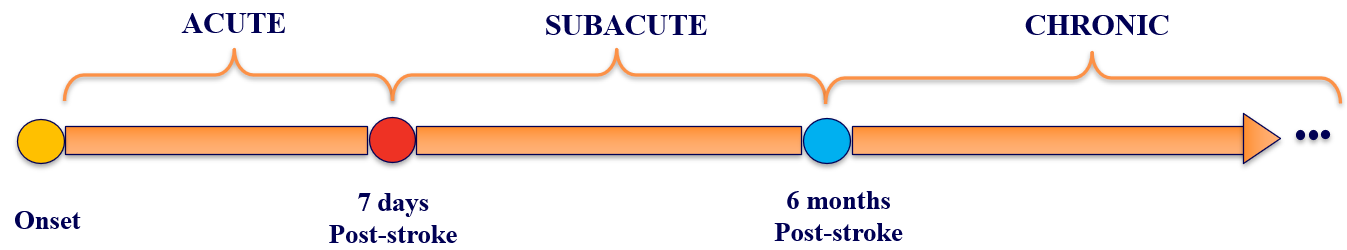}
    \caption{The phase of post-stroke recovery stages \label{Fig:term1}}
\end{figure}

Throughout this review, the terms: acute, subacute and chronic stages refer to the three phases of recovery after stroke. The timeline of these phases, which spans from the time of the initial stroke to years afterward, is summarized in Figure \ref{Fig:term1}. Their definition is in accordance with the recommendation of~\cite{RN216} and previous studies in stroke rehabilitation programs~\cite{RN218, RN217}. 

\begin{figure}[!ht]
    \includegraphics[width =0.85\textwidth]{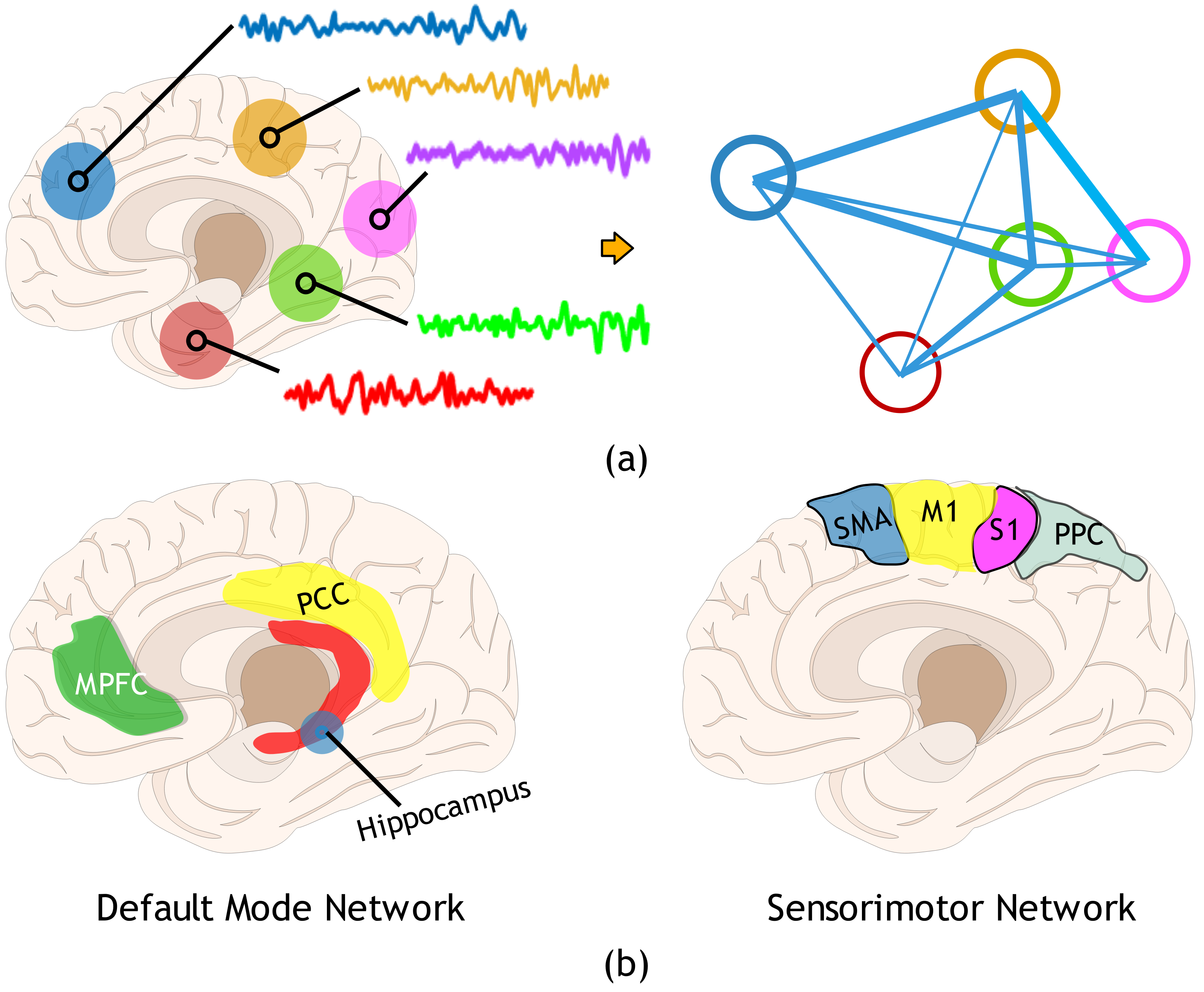}
    \centering
    \caption{ \textbf{(a)}. Inferring functional connectivity among multiple brain regions by calculating pairwise correlation; FCN: a functional connectivity graph can be generated where the vertices represent the brain regions and the edges represent the strength of FC/FNC between these regions. \textbf{(b)} The large-scale functional networks: the default mode network (DMN), which involves the medial prefrontal cortex (MPFC), posterior cingutate( PCC), bilateral hippocampus; and the sensorimotor network (SMN), which includes supplementary motor areas (SMA), primary motor cortex (M1), primary sensory area(S1), posterior parietal cortex (PPC). \label{Fig:FNC}}
\end{figure}

\subsubsection*{Functional Connectivity(FC), Functional Network Connectivity(FNC) and Functional Connectivity Network(FCN)}

The terms Functional Connectivity (FC), Functional Network Connectivity (FNC), and Functional Connectivity Network (FCN) are frequently used by the authors in the studies reviewed in this paper. These nouns are so similar that the readers risk becoming confused if they are not attentive.
Here, FC is defined as correlation (or other statistical dependencies) among spatially remote brain regions~\cite{RN220}. The process of inferring functional connectivity among multiple brain regions by calculating pairwise correlation is summarized in Figure \ref{Fig:FNC}\textbf{(a)}. FNC can be seen as a higher level FC, which refers to a statistical dependency among large-scale functional networks (or functional domains) in the brain~\cite{RN221}, for example, the default mode network(DMN)~\cite{RN223}and sensorimotor network (SMN)~\cite{RN225}. A representation of brain regions that the involved in functional networks may be found in Figure \ref{Fig:FNC}\textbf{(b)}. If with no specific statement, the FC and FNC refer to a pairwise Person’s correlation in this review. By contrast, Functional Connectivity Network (FCN) is a concept based on FC and FNC. It refers to a functional connectivity graph where the vertices represent the brain regions and the edges represent the strength of FC/FNC between these regions.

\subsubsection*{Motor recovery measurement}

Clinically, motor recovery after stroke is measured by the absolute difference between baseline and subsequent motor function scores~\cite{RN230}. This can be best achieved by comparing the motor function assessment at longitudinal time points. In the studies reviewed in this paper, there are several methods used by the authors to evaluate the post-stroke motor function, including the(FMA)scale~\cite{RN231}, which evaluates patients' single-joint and multi-joint motor ability, loss of co-motor energy, finger individualization ability, movement speed, measurement impairment, ataxia, and motor reflex, and the Paralysed Hand Function Assessment Scale~\cite{RN232}which measures the degree of paralysed hand for stroke patients. Other evaluation methods include the Action Research Arm Test (ARAT)~\cite{RN227} Modified Rankin Score(mRC)~\cite{RN228}. etc. 

Beyond that, comparing the post-stroke recovery stages that the stroke patients in  at different time points can helps demonstrate the motor recovery process. One of the most well-known methods of measuring the stroke recovery stages is the Brunnstrom stages, also known as the Brunnstrom approach~\cite{RN229}. By classifying the assessment score into distinct zones, the National Institutes of Health Stroke Scale (NIHSS) or the evaluation scale indicated above are widely used to estimate the patient's stroke recovery stages.
 This partition can also be utilized to divide the patient sample into subgroups affected by stroke. Hence, in many cross-sectional studies, the relationship between the FC and motor recovery can be investigated by examining the FC alteration between different groups~\cite{RN234,RN235,RN236,RN233}

\section*{fMRI processing for DFC and SFC }
\subsection*{Preprocessing and denoising RS-fMRI}
By sampling the brain's three-dimensional (3D) volume every 1-2 s (or faster), MRI scanning can obtain the brain landscape map at the millimetric spatial resolution. Then, the BOLD signal contrast can be extracted within a full MRI scan. The BOLD signal variance can represent the resulting brain neural activity through multivariate time series. As the BOLD signal is weak and suffers from the noise of multiple sources, the raw fMRI data needs to undergo extensive pre-processing before further analysis.
Pre-processing of rs-fMRI signals included the following steps: 
\begin{enumerate}
 \item  removing a number of volume from the beginning for steady BOLD signal (typically, 3~\cite{RN202} or 10~\cite{RN62});
\item  slice timing correction
\item realignment for head motion
\item outlier detection for scrubbing;
\item registration to structural images, segmentation~~\cite{RN60} and lesion-masked normalization~~\cite{RN61};
\item spatial smoothing using a full-width at half-maximum (FWHM) Gaussian kernel (4 or 8 mm for the recommendation).
\end{enumerate}
Note that this is a general possible pre-processing pipeline. The processed methods and their orders for the studies can be varied across specific applications and tasks. 

\begin{figure}[!ht]
    \includegraphics[width= 0.95\textwidth]{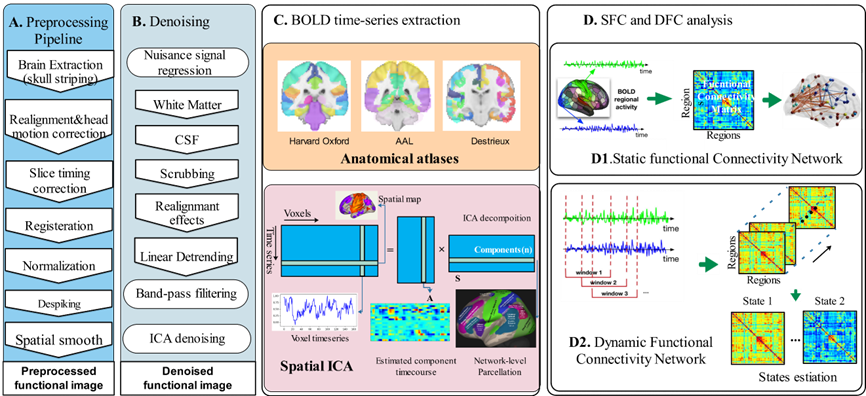}
    \caption{The pipeline of fMRI processing for SFC and DFC analysis.\textbf{A}.Preprocessing functional images \textbf{B}. Denoising functional images. \textbf{C}.Atlases based and Spatial ICA based time series extraction \textbf{D1}. SFC analysis. \textbf{D2}. Sliding window based DFC analysis\label{Fig:pipe}}
\end{figure}

Despite the pre-processing processes' capacity to remove the majority of brain activity disturbances, the BOLD signal often still contains considerable noise or non-neural variability due to a combination of a physiological, outlier, and residual subject-motion effects~\cite{RN59,de2020noise,ciric2017benchmarking}. These residual noise components in the BOLD signal will introduce strong and observable biases in all functional connectivity measures. Therefore, there are typically additional strategies to remove or at least minimize these underlying interferences in the context of functional connectivity. These strategies generally refer to denoising step, which include linear regression of potential confounding effects  (e.g. effect from the gray matter, white matter, and cerebrospinal fluid) in the BOLD signal, linear detrending, temporal band-pass filtering. Similarly, the denoising strategies has not gold standards, they are variable across the studying methods and tasks. Figure \ref{Fig:pipe} \textbf{A} and \textbf{B} exhibit various preprocessing steps and denoising strategies. One note that nearly all studies reviewed in this paper introduce the preprocessing step, while the denoising step has been not be emphasised. The denoising step has been proved that it can benefit FC estimation by improving the signal quality and reducing the motion artefacts~\cite{de2020noise}. Hence, these additional denoising strategies should be encouraged to gradually introduce into stroke studies, which probably can enlarge the variability in FC alteration caused by stroke and then promote more reliable and accurate results. 

Several existing open-source tools can perform the pre-processing and denoising steps and enable the analyzable fMRI to be obtained reliably and rapidly, including the Statistical Parametric Mapping software package (SPM), the FMRIB Software Library (FSL)~\cite{RN58}, the Data Processing Assistant for Resting-State fMRI (DPARSF)~\cite{RN57}, CONN toolbox~\cite{RN59}, the graph-theoretical network analysis toolbox (GRETNA)~\cite{RN191}. 

\subsection*{Time series extraction}
After the raw RS-fMRI has been pre-processed and denoised, the time course of the brain areas during the fMRI scan has to be extracted for further post-stroke functional connectivity analysis. The methods for time series extraction are mainly divided into two categories: ROI-based and data-driven methods.

The ROI-based method relies on predefined regions of interest (ROI) to extract the time course of brain areas. As the motor control region in the brain system, nearly all investigations linking to post-stroke motor active recovery covert the primary motor cortex (M1) as one of their ROIs. Meanwhile, Supplementary Motor Area (SMA) is frequently investigated as an ROI. The decreased connectivity between M1 and SMA has been found across studies in stroke patients. Beyond M1 and SMA, other regions, including the premotor cortex, and inferior frontal gyrus, are also predefined as ROI to investigate the inter-regional abnormal connectivity following stroke. In addition to manually set regions, the anatomical atlas is often used to identify ROIs. There are typically three anatomic atlases utilized to define brain regions: Destrieux~\cite{RN64}, Harvard-Oxford~\cite{RN65} and AAL~\cite{RN66} atlas(see Figure \ref{Fig:pipe} \textbf{C} for the three anatomical atlas).  

Data-driven method, by contrast, does not need to pick brain seeds beforehand, and it is free from prior knowledge of specific brain regions. Specifically, it uses multivariate voxel-wise projection techniques such as independent component analysis (ICA) or its variants (e.g., group ICA) to decompose the raw fMRI data into multiple independent components (ICs). Each IC represents the brain areas with independent brain activity. Figure \ref{Fig:pipe} \textbf{C} shows the diagram of the spatial ICA based timeseries extraction method. Then, these ICs will be visually inspected to identify if the brain areas they represent can have brain activity because of a nerve action or if this activity is just caused by noise.The spatial map of IC with true brain activity can be matched with the previous anatomic brain function map.Meanwhile, their power-frequency curve has a specific fluctuating pattern(peak at low frequency, decrease rapidly, and then remain relatively stable). The identified ICs were recognized as intrinsic connectivity networks(ICN) which can be utilized for further functional connectivity analysis between functional areas or high-level functional connectivity between brain functional domains(e.g., the default mode, sensorimotor, subcortical, and cerebellar functional network).

\subsection*{Static and dynamic functional connectivity estimation}
\subsubsection*{Static functional connectivity estimation}
Typically, SFC estimates functional connectivity between two brain regions by calculating the pairwise correlations between their time series. In the case of fMRI analysis, this correlation is frequently quantified with a canonical correlation coefficient~\cite{RN69, RN70, RN38} that measures the similarity of amplitudes of BLOD fluctuations. There are also studies using the indices-based phase coupling measure, including the coherence~\cite{RN67} or phase-locking relationships~\cite{RN68}. The pairwise correlations between region nodes or functional domains can construct a Functional Connectivity Network (FCN) representing the brain's functional connectivity profile. Usually, before performing the group-level FC analysis, the individual correlation-based FC is often transformed to a normalized distributed Z-score using Fisher’s r-to-Z transformation, as it will not be bounded by upper or lower limits, which is helpful for further statistical analysis.    

Based on the pairwise correlations, many SFC analysis methods exist, such as the FCN density analysis method~\cite{RN77}, the Regional homogeneity (ReHo) based approach~\cite{RN213} and Kendall correlation coefficient (KCC) approach~\cite{RN222}. In addition, considering the investigated brain regions as nodes and the pairwise correlations as edges, the functional connectome can be viewed as an adjacency matrix or graph. Hence, the FCN can be naturally analysed with graph-based methods. For example, the clustering coefficients, global efficiency and local efficiency of the FNC graph can be calculated with graph theory~\cite{RN214}, and these graph-based measures can be integrated to analyze the functional connectivity of the brain.

\subsubsection*{Dynamic functional connectivity estimation}

The development of DFC analysis for RS-fMRI is incited by the fact that the FC/FNC between regions or voxels may change in a short period~\cite{RN71}. Accumulating evidence has demonstrated that the FC/FNC alternation follows certain functional coupling patterns over time~\cite{RN73, RN72}, and these fluctuating coupling patterns support the brain regions in processing different functional requirements~\cite{RN72}.

A multitude of approaches to quantify these dynamic patterns and their properties have been introduced (the detailed review related to dynamic functional connectivity can be seen in~\cite{RN74, RN72,RN75, RN47}). The most common and straightforward way to measure the DFC is by calculating the FC/FNC between regions/voxels in a consecutive window period. The window length is usually constant, and the sliding scheme spanning the full scan is run. The length of the sliding step is generally less than the window length; thus, two adjacent windows can partially overlap. With the sliding window method, thousands of FC/FNC matrices can be generated at the window length level. These successive matrices exhibit the dynamic evolution of brain state within an MRI scan duration, and thus, the fluctuations in the connectivity time courses can be easily assessed. The measurements to quantify the fluctuation include the temporal standard deviation~\cite{RN49, RN117, RN105}, coefficient of variation~\cite{RN110}, or amplitude of low-frequency fluctuations (ALFF)~\cite{RN109, RN108}. In addition, the matrix factorisation techniques can summarise a series of FNC patterns into multiple FNC states. The recurrent or repeating connectivity patterns/states represent the underlying functional coupling patterns. The typical method to summarize the connectivity patterns is K-means clustering~\cite{RN111}, which has been widely used in fMRI studies for DFC analysis since a study by Allen et al~\cite{RN98}. The clustering approach reveals the reoccurring connectivity patterns from the windowed FC matrices, and the dynamic brain states corresponding to ongoing processing can be reflected. 

Another representative approach summarising the brain state is the innovation-driven co-activation patterns (iCAPs) method~\cite{RN75}. iCAPs method does not rely on the sliding window scheme but needs a clustering algorithm's assistance. It identifies BOLD signals' transients (or innovation frames) and feeds them into clustering methods to obtain iCAPs~\cite{RN47, RN113}. Besides, there are also other DFC methods, such as the principal component analysis (PCA) method~\cite{RN100}, the dictionary learning (DL) method~\cite{RN101, RN99}, the tensor decomposition(TE) methods~\cite{RN104}, the dynamic community detection based method~\cite{RN102, RN103}, the Hidden Markov Model(HMM)~\cite{vidaurre2017brain} method and the wavelet transform coherence (WTC) method~\cite{RN238}. However, in studies about post-stroke motor dysfunction, these methods are rarely investigated for DFC analysis. 

\subsubsection*{SFC and DFC comparison}

In principle, the SFC method assumes that FC/FNC is stationary over MRI scan time, and the functional connectivity is constant during a MRI scan. Nevertheless, SFC ignores that the human brain is a dynamic system that fluctuates even at the time scale of milliseconds~\cite{RN215}. Therefore, DFC analysis approaches developed rapidly in the last decade to investigate the dynamic nature of the brain. In methodology, the SFC analysis approaches are just based on the average FC/FNC that aggregates over an entire fMRI scan. Despite SFC’s methodological simplicity and ease, SFC approaches can vividly show the disease-induced FC alterations. This is particularly important when investigating the damage's impact on the specific brain functional area. For example, many post-stroke studies have employed SFC to investigate the stroke-lesion caused FC changes and to verify if the post-stroke FC reorganization can underpin functional recovery~\cite{RN234,RN235,RN236,RN233,RN237,RN36,RN44}. For that reason, the SFC can provide a valuable reference to functional rehabilitation programs. Many disease treatments have utilized invasive or non-invasive tools (e.g. the transcranial magnetic stimulation~\cite{RN34,kadono2021repetitive}) in clinical trials to help the affected brain motor function areas promote recovery~\cite{RN219}. 

By contrast, DFC analysis approaches employ time-resolving or signal decomposition methods to investigate a time-varying FC/FNC that span the fMRI scan. It can extract relevant dynamic features to reflect functional network flexibility and brain state transition. However, despite this potential of dynamic FC analysis, only a handful of studies have employed such time-resolved approaches to explore the dynamic neural mechanism following stroke. In contrast to the thrive research of SFC methods in other neurological diseases such as Parkinson’s disease~\cite{RN114} and Huntington’s disease~\cite{RN116}, DFC analysis on rs-fMRI following stroke is still developing. Besides, compared with SFC analysis, DFC approaches typically have a complicated mathematical/probability theories~\cite{RN230}(e.g., the HMM-based DFC analysis~\cite{vidaurre2017brain}). Hence, DFC is probably not friendly to those clinicians without mathematical or engineering backgrounds. Furthermore, due to increased fMRI temporal-spatial resolution, the DFC is more sensitive to noise~\cite{RN72,RN47}. Thus, the extra hype-parameter setup (e.g., the window length and sliding step in the sliding-window DFC method) needs to be fine-tuned to verify the validity of the results~\cite{RN74, RN72}. 

Overall, SFC and DFC both have advantages and disadvantages(summarized in Table \ref{tab:comparsion}). Compared with DFC, SFC analysis is more mature and has more applications in post-stroke fMRI analysis; however, the DFC analysis can capture the intrinsic dynamic nature of the brain, which shows a potential promising prospect. Even though DFC models can rarely be seen in post-stroke studies currently, they are expected to boom in the future of post-stroke investigation.


                                      

\begin{table}[]
\centering
\caption{SFC and DFC comparison.\label{tab:comparsion}}
\renewcommand\arraystretch{1.5}
\begin{tabular}{|c|p{1.8cm}|p{3cm}|p{3cm}|p{2.3cm}|}
\hline
- & Definition & Methods        & Pros        & Cons  \\ \hline
SFC &  \begin{tabular}[c]{@{}l@{}}FC   keeps static \\in fMRI scan \end{tabular}& \begin{tabular}[c]{@{}l@{}}Pairwise   correlation;\\  FCN density~\cite{RN77};\\ Regional homogeneity\cite{RN213};\\Graph theory~\cite{RN214};\end{tabular}   & \begin{tabular}[c]{@{}l@{}} Mature application;\\ Brian region focused;\\ Vivid FC alteration;\\ Applied in clinical trials.\end{tabular} & \begin{tabular}[c]{@{}l@{}}Methodological \\ simplicity and ease;\\ Ignore   brain\\ FC dynamics.\end{tabular}  \\ \hline

DFC &\begin{tabular}[c]{@{}l@{}} FC   continuously\\ flactuates\end{tabular}  & \begin{tabular}[c]{@{}l@{}}Sliding   window~\cite{RN98};\\ PCA method~\cite{RN100};\\  DL method~\cite{RN101, RN99} \\ TE method~\cite{RN104};\\ DCD method~\cite{RN102, RN103};\\  HMM~\cite{vidaurre2017brain};\\ WTC~\cite{RN38} ;\end{tabular} & \begin{tabular}[c]{@{}l@{}}Increased temporal-spatial\\resolution;\\Reflective  network \\flexibility;\\Exhibitaion of brain\\state transitions\end{tabular} & \begin{tabular}[c]{@{}l@{}} Complicated \\mathematics~\cite{RN230};\\Sensitive to noise;\end{tabular}                 \\  \hline
                                      
\end{tabular}
\end{table}

\section*{Static connectivity analysis}
\label{sec:SFC}
\subsection*{Decreased functional connectivity is a common finding}

In numerous rehabilitation studies estimating resting-state FC, a decrease in interhemispheric FC was commonly found in the functional network or at least between some regions by comparing results from stroke patients with healthy controls. This reduction is caused by the disruptions in multiple large-scale functional brain networks, which has been viewed as one of the characteristics of motor impairment following stroke~\cite{RN44}. 

In detail, the lowest level of FC appears in the acute phase of stroke, and the earliest time of this finding is within a few hours post-stroke. Hence, interhemispheric FC reduction is often reported in many studies studying FC in acute ischemic stroke patients. For example, Liu et al.~\cite{RN77} observed disrupted interhemispheric FC between the motor cortices of acute stroke patients, which was associated with motor deficits. A recent study investigating dynamic structural and functional reorganizations following motor stroke reported a significantly lower interhemispheric FC in the first week~\cite{RN55}. This decrease in interhemispheric connectivity is more distinct than those in intrahemispheric connectivity. In addition, Nai-Fang et al~\cite{RN51} observed decreased interhemispheric FC within the cortical motor network in the first acute unilateral ischemic stroke patients. Specifically, the FC between the ipsilesional M1 and contralesional M1, contralesional postcentral gyrus (PoCG), dorsolateral PMC are significantly lower than that in controls. Besides, the interhemispheric FC reduction can also be observed in the post-stroke subacute and chronic stages. In~\cite{RN36}, the FC between M1 and the contralateral cerebral cortex was reduced in stroke patients with unilateral ischemic motor neural network injury 2 weeks after onsets. And the stroke participants recruited by~\cite{RN62} showed a decreased FC between the ipsilesional SMN and the contralesional SMN and auditory network (AN), which supports the early findings in the~\cite{RN78} that the disruption of interhemispheric interactions between bilateral SMNs may result in motor deficits in patients with chronic stroke.

Beyond interhemispheric FC, the decreased contralesional M1 FC can be found in a study~\cite{RN79}. The lowest FC levels in the contralateral sensorimotor cortex are 2 weeks after stroke. The~\cite{RN210} reported the reduced within-network FC in the contralesional precentral gyrus within the dorsal sensorimotor network and the contralesional superior parietal lobe.

\subsection*{Brain functional connectivity network topology supports the decreased FC}

Brian functional connectivity network (FCN) can be described as a set of nodes (ROIs or independent components ) connected with edges (functional connectivity measure). Thus, graph theory analysis on FCN can provide important information about the topological properties of the brain functional network. When analyzing FC from the perspective of a brain graph, the various FC network topology measures also approve of the reductions in FC in contrast to a healthy brain. 

For the local topological measure of the FC graph, a study has proved the shortest path length to be lower than healthy controls in acute ischemic stroke~\cite{RN50}. And the clustering coefficient was reported to decrease~\cite{RN85,RN84, RN81}, though~\cite{RN50} suggested that stroke patients maintained the local clustering coefficient. In addition, another local measure, the weighted node degree which measure the number of connections has been found to decrease in the contralesional M1 of patients with bad recovery post-stroke~\cite{RN80}.There are no significant finding in topological measures between patient and control was reported in other studies~\cite{RN81} following stroke.

Regarding the global topological measure of FCN, a reduction in network modularity was observed in an experiment including 25 patients with focal lesions because of stroke~\cite{RN82}. This finding was consistent with the decrease in interhemispheric function integration. Modularity measured by the connection density within communities over that between communities reflects the degree of functional integration and segregation~\cite{RN83}. A recent study supported the findings in~\cite{RN53}, where significantly lower modularity was found in patients compared with controls, indicating decreased segregation in the FNC. Beyond modularity, other global measures of the FC network tended to decline. In~\cite{RN81}, the FC concordance, which measures the network stability in time, was observed to decrease over time in contrast to intact networks. In addition, the small-worldness which reflects the ability of brain networks satisfying the needs of local and global information processing was significantly lower in patients than controls two weeks after stroke~\cite{RN211}. 
	
There are a great varity of graph theory based topology measures, resulting in the dynamics patterns in the network topology investigated in studies are various. In general, the graph topology alteration of FCN post-stroke can be interpreted in a concept of network randomization~\cite{RN82}, the way the network reorganizes itself to adapt to the lost function, which also demonstrates the process of neuroplasticity that occurs in the brain post-stroke.

\subsection*{Increased functional connectivity is positively related to recovery}

Even though a decreased FC is a dominant trend in patients following a stroke in contrast with healthy controls, many findings also include the increased FC. Several longitudinal studies have demonstrated that the interhemispheric FC in the SMN first decreases in the early stages after stroke while increases in the following weeks or months~\cite{RN91, RN33, RN79}. Other studies support this FC-increased finding as well. For example, in a systemic study~\cite{RN44}, the decreased FC was observed in stroke patients between the ipsilesional M1 and the sensorimotor cortex, the occipital cortex, the middle frontal gyrus(MFG) and the posterior parietal cortex, while the increased FC was also shown between ipsilesional M1 and cerebellum, the thalamus, the middle frontal gyrus (MFC) and the posterior parietal cortex. Besides, in the study~\cite{RN77}, although the FC in the motor area decreased after stroke, the opposite occurred in cognitive networks. In addition, a recent study~\cite{RN49} reported that compared with healthy groups, the patients exhibit a significantly increased static FC between a large number of structures, including ipsilesional M1 and the contralesional precentral gyrus(PrG), contralesional M1 and the ipsilesional PrG, contralesional precuneus;  ipsilesional MFG and precuneus, the contralesional cerebellar, PoCG,  ipsilesional sub-gyral region; SMA and ipsilesional PrG, frontal-temporal space, MFG; ipsilesional  SMA and the ipsilesional middle temporal gyrus (MTG).

The increased FC post-stroke seems to be restricted between specific structures~\cite{RN23}, while the highly consistent finding is that the increased FC occurred in connectivity with the cerebellum~\cite{RN44, RN81, RN95}. This conclusion derived from a potential recovery mechanism that FC in the cerebellum was reported to be crucial for recovery~\cite{RN89, RN88}. In fact, not limited to the cerebellum, many structures that found increased FC correlate with motor recovery. The commonly mentioned brain region is M1. Many studies have reported that FC increases between the ipsilesional M1 and contralesional M1 and between the ipsilesional SMC and contralesional SMC,which is related to better recovery~\cite{RN77, RN97, RN95}. Besides, Ktena et al. in~\cite{RN86} observed increased FC between lesion areas and M1 in the unaffected hemisphere of the chronic patient. They concluded that this phenomenon reveals the network reorganization process associated with motor recovery. Furthermore, the enhanced FC between M1 and SMA and other motion-related regions like the dorsolateral prefrontal cortex (PMD) has been be archived and proved that this enhancement is a potential mechanism for motor function recovery~\cite{RN86, RN90}. 

\begin{sidewaystable}[]
\centering
\caption{The summary of studies using post-stroke rs-fMRI based static function connectivity for motor network analysis.}
\begin{tabular}{|C{0.14cm}|C{1cm}|c|C{1.5cm}|C{0.5cm}|C{1.5cm}|C{1.2cm}|p{2cm}|C{4cm}|C{6cm}|}

\hline
 & Studies & Sub.	& Stoke Info. &	Ctrl. & Type 
& Timing &	Func.Assessment.	& sFC Method	& Findings\\ \hline


1 &~\cite{RN55} & 34 & unilateral ischemic stroke
& 34 & Longitudinal & 1w, 4w, and 12w 
& NIHSS; FMA.	& Seed-based: ROIs selected in somatomotor network (SMN) along the left and right primary motor cortex (M1)	& A significantly lower interhemispheric FC in the first week and a continual increase to week 12; \\ \hline
 
2 &~\cite{RN51} & 67 & acute unilateral ischemic stoke & 25 & Cross-sectional & Within 7d & NIHSS;
mRS(90d) & Seed-based: predefined ROIs: primary motor cortex (M1), supplementary motor area (SMA), PoCG, ventrolateral PMC, and dorsolateral PMC, posterior cingulate cortex, anterior medial prefrontal cortex , inferior parietal lobule , retrosplenial cortex, and anterior hippocampus.
& Significantly lower FC between iM1\&cM1,iM1\&cPCG,iM1\&cPMd,
iPCG\&cM1, iPCG\&cPoCG, iP0CG\&cPMd, and cM1\&cPCG; FC of iM11\&cM1, iM11\&cPoCG, iPCG1\&cM1,iM1\&cPMC was significantly
lower in patients with unfavorable outcomes than in patients with favorable outcomes. \\ \hline

3 &~\cite{RN62} & 52 & chronic subcortical stroke & 52 & Cross-sectional & $>$ 3m & FMA & ICA based:15 networks & Decreased FC between the ipsilesional SMN and the contralesional SMN and AN. \\ \hline
    
4 &~\cite{RN36} &
81& unilateral stroke & 55 & Longitudinal &	2w,3m	& FMA &	Seed-based: Predefined 24 ROIs &	FC were weakened 2 weeks post-stroke and strengthened for 3 months. \\ \hline

5 &~\cite{RN49} & 76 & unilateral stroke &	55	& Cross-sectional & from 7d to 1m &	NIHSS;
FMA &	Seed-based: predefined 6 ROIs( left M1, right M1, left PMC, right PMC, left SMA, and right SMA).	& Increased FC between cM1\&cPrG,iM1\&iPrG, precuneus\& cPMC, iMFG\&Precunesu, cCerebellar\&cPoG, iPMC\&cPrG, \&cIPL\&I, cSMA\&iPrG, cIFG\&cMFG, iSMA\&iMTG;
No significant between sFC and clinical measure. \\ \hline

6 &~\cite{RN52} &
37 & ischemic stroke &	-  & Cross-sectional &
3d;90d &	NIHSS(at 3d);
mRS(90d) &	Atla-based: Harvard-Oxford &	Good functional outcome had greater functional connectivity right temporal lobe and left frontal lobe, between the left temporal lobe and right frontal lobe and between the right temporal and parietal lobes; \\ \hline

7 &~\cite{RN86} &
41 & ischemic stroke &	-	& Cross-sectional &
2-5d; 	& NIHSS;
mRS(90d) &	Atla-based:three atlas: Destrieux, Harvard-Oxford and AAL atlas &	The worse outcome showed higher values of characteristic path-length. \\ \hline

8 &~\cite{RN54} & 24 &	chronic supratentorial stroke &	- & Cross-sectional & $>$2M &	FMA	& ICA based:16 regions: supplementary motor area (SMA); left dorsolateral
prefrontal cortex (DLPFC); right frontoparietal network; left frontoparietal network; 
cerebellum; 	& The internetwork connectivity was significantly increased in the mild group for SMA–M1 in the affected hemisphere and SMA–dorsolateral prefrontal cortex DLPFC in the unaffected hemisphere and for lesion–M1 in the unaffected hemisphere compared with the severe group;Increased internetwork connectivity between remote brain regions may result in the reorganization associated with motor recovery.\\ \hline

9 &~\cite{RN209} & 42 &	chronic &	- &	Cross-sectional & $>$2w &	NIHSS;FMA & Atla-based; & Whole-brain	Interhemispheric homotopic functional connectivity, correlated with
improvements in Upper-extremity function. \\ \hline

10 &~\cite{RN53} & 65 &	Stroke & 25	& Longitudinal & 2w;3m;1y &	Action Research Arm Test;
Functional Independence Measures walk test; &	Atla-based:324 regions of interest & The degree of integration within networks and segregation between networks were significantly reduced sub-acutely but partially recovered by 3 months and 1 year; network recovery not correlated with recovery from motor. \\\hline

11 &~\cite{RN77} & 8 & Stroke & 10 & Longitudinal &1w; 2w; 1m; 3m; 1y	& Motricity Index;
NIHSS	& Seed-based: ROIs in sensorimotor regions (SMC, inferior parietal lobule) and non-sensorimotor regions(DLPFC)	& Between the contralesional SMC and ipsilesional SMC, FC  decreased in the first 2 weeks and followed by increases towards normal levels.\\\hline

\end{tabular}
\end{sidewaystable}

In many studies related to the prediction of motor recovery or outcomes, the increased FC is normally associated with minor severity or a better motor functional outcome. In a study that included 34 patients and healthy control, the author found a significantly lower interhemispheric FC in stroke patients compared to healthy controls in the first week, which support the previous common finding. However, after that, the FC continually increased to week 12. The correlation analysis shows that the percentage of FC changes was significantly positive with the improved FMA score from week 1 to week 4~\cite{RN55}. In an early prediction study~\cite{RN52}, 37 stroke patients were scanned on day 3 after stroke, and the fMRI data were used to predict 90-day outcomes. The results show that patients with good outcomes had higher FC than those with poor outcomes. Adding the FC improves the model's accuracy to 94.7\%, reflecting that increased FC plays an essential role in motor recovery. The stroke patients at the chronic stage have a similar increase. In a study with a total of 107 participants, compared to the patients with a completely paralyzed hand, the patients with a partially paralyzed hand had increased FC in the ipsilesional superior temporal gyrus, the ipsilesional middle occipital gyrus and the contralesional calcarine~\cite{RN62}. This finding is also enhanced in a larger cohort of patients with ischemic stroke at the acute stage. In a study with 85 stoke acute ischemic patients, the FC between ipsilesional M1 and contralesional PMD in patients with favorable outcomes was significantly greater than with unfavourable outcomes~\cite{RN51}, which demonstrate that the increased FC can serve as an independent outcome predictor.

In total, the FC of the motor network is impaired after stroke onset, and the decreased FC is a consistent finding, but the increased FC can be observed. The increased FC is a way of neuroplasticity, which means that the lost FC has tended to grow to the normal level. From the perspective of neural activation, the decreased FC implies the existing pathways are disinhibited in the recruitment stage after stroke~\cite{RN21}, and the increased FC shows that the damaged pathways surrounding the lesion area are newly built thus improving motor function.

\section*{Dynamic connectivity analysis}
\label{sec:DDFC}
For patients with motor dysfunction following stroke, the number of good examples of DFC analysis is not as high as SFC. However, from the existing investigations, we can conclude the following findings.

\subsection*{The temporal variability of FC is altered following stroke.} 

DFC analysis methods resort to the temporal variability of FC to reveal neural dynamic properties and recovery mechanisms of stroke. The temporal variability between specific regions illustrates the dynamic reconfiguration of the brain system over time in response to going processing and globally reflects the degree of synchronization between functional areas in the brain. The definitions of temporal variability are variable across studies.

 The study~\cite{RN117} calculated the temporal variability between specific regions as the average functional connectivity over different windows. By contrast, the temporal variability of FC in~\cite{RN49} was characterized as the standard deviation of time courses at predefined seed regions across a series of windows. In terms of the difference, Hu et al. in~\cite{RN117} reported a significantly reduced temporal variability in stroke patients compared to the healthy group; however, the FC temporal variability decreased-regions exhibited disparity at distinct poststroke stages. In the acute stage, the reduced regions cover the primary sensorimotor and default mode network (DMN), while only the ipsilesional PoCG and ipsilesional anterior cingulate gyri (ACG) showed a declining trend in the subacute stage. Nevertheless, this finding is incompatible with a study~\cite{RN49} finding, where an increased temporal variability in ipsilesional M1 and contralesional PrG was observed. And this increase shows a longitudinal trend, which is exhibited over the stages of stroke. 
 
Taking advantage of the temporal variability, the relationship between dynamic FC and motor function recovery poststroke was investigated in both studies.  In~\cite{RN49}, the authors detected a significantly negative correlation between FMA scores and FC variability in ipsilesional M1 and contralesional PrG. And in~\cite{RN117},  the increased FC variability from the acute to the subacute stage was reported to correlate positively with the increased FMA.  

The results across studies appear to be different, or to an extent, the opposite. This exhibits a preference for these publications. It is not expected but can be accepted as the subjects, the stroke severity, the post-stroke stage, the brain motor area, and the algorithm calculating temporal variability are biased in various investigations.

\begin{table}[]
\centering
\caption{The summary of DFC analysis with fMRI for post-stroke recovery.}
\resizebox{\textwidth}{!} {

\begin{tabular}{|c|l|p{1cm}|l|c|p{1.2cm}|p{1.8cm}|p{2cm}|p{5cm}|}
\hline
Studies & Sub.	& Stoke Info. &	Ctrl. & Type 
& Timing &	Extraction methods.	& DFC Method	& Findings\\ \hline

~\cite{RN49} & 19 &
Ischemic stroke & 19 & Longitudinal & 7d/2w/3m &	Seed-based: AAL 116 ROIs & Sliding window approach; & FC temporal variability:	Reduced temporal variability; Longitudinal increased over the stages. \\ \hline

~\cite{RN117} & 75 & Ischemic stroke & 55 & Longitudinal  &	7d-1m &	Seed-based: six ROIs (bilateral M1, SMA and PMC ) &	sliding window approach; standard deviation & Increased dynamic FC between the ipsilesional M1 regions and contralesional PrG, and a negative correlation between dFC in the regions and FMA scores after stroke. \\ \hline

~\cite{RN119} & 31 & Ischemic stroke  & 17 & Longitudinal & within 2 week & Voxel-based: spatially constrained ICA:13 network components	& sliding window approach; k-means clustering	& Severe subgroup:  spatially segregated connectivity configuration; regionally densely connected. increased transition likelihood to this State 1
Moderate: weakly connected configuration (low levels of connectivity) spent more time. \\ \hline

~\cite{RN212} & 41 & Ischemic stroke & - & Cross-sectional & $<$7d;$>$6m & Voxel-based: ICA (49 to 7 functional domain) &	sliding window approach; k-means clustering. &	NIHSS significantly correlated with fraction and dwell time of state 1. \\ \hline
 	
~\cite{RN63}  & 54 & Ischemic stroke & - & Cross-sectional &	$<$7d;$>$6m & Voxel-based: ICA(14 to 3 functional domain) &	sliding window approach; sparse inverse covariance matrix; k-means clustering. &	Mildly: lower variability values Moderately-to-severely: higher dynamic connectivity variability fraction and dwell time improve the prediction performance. \\ \hline

\end{tabular}
}
\end{table}

\subsection*{Motor function affected patients alters the preference in transient brain states of sMN}

The mutual transition of connectivity state in diseases with highly dynamic neural activity abnormalities has been active research in DFC analysis. The so-called connectivity state is generally abstracted from the reoccurring BOLD signal using a sliding window scheme and clustering algorithm. Thus, a successive highly recurrent FC pattern list within an MRI scan can vividly exhibit a dynamic transition between the multiple brain states. This dynamic transition globally reflects the FCN's flexibility and can be utilized to investive the alteration or adaptation of dynamic interaction between brain functional networks after stroke.

The alteration of connectivity states’ flexibility and adaptation due to cognition and psychiatric disorder has been illustrated in other studies, while stroke-induced changes in brain states have rarely been investigated. Recently, the abnormal connectivity states in acute ischemic stroke (AIS) attracted the attention of researchers. 

In a study which included 31 AIS patients~\cite{RN119}, the authors outlined three different SMN connectivity states: the first is characterized by extremely strong intra-domain connectivity and extremely weak inter-domain connectivity; the second has remarkably weak intra-domain connectivity; and the third is a compound state that combines the characteristics of states 1 and 2. The summarized states do not differ too much from the states summarized in their further work~\cite{RN63}\cite{RN212}, while three studies revealed different aspects concerning post-stroke motor impairment.~\cite{RN119} provides the distinct configuration of the FC connectivity states in stroke patients with various degrees of clinical symptoms. Moderately affected patients, for example, have significantly more dwell time in a weakly connected configuration, while severely affected patients prefer to stay in the state 1. This finding is consistent with the study~\cite{RN63}, which included 41 AIS patients. This study also demonstrated that NIHSS significantly correlated with a fraction (the ratio of the time a subject spent in a given state over the scan time) and dwell time  (the time a subject spent in a state without switching to another one ) of state 1. \cite{RN212} has the most AIS patients (54) among the three studies. In this study the author pays more attention to distinguish the link between the SMN connectivity measure and the subgroups with or without the motor deficit. Results show that embedding the fraction and dwell time into the initial motor impairment-based model can improve the prediction performance(95\% accuracy). 

Note that this finding does not derive from multiple independent investigations. Hence, it needs to be validated in other stroke patient cohorts to test if the results are reliable and reproducible. Additionally, only the SMN has so far shown the preference shift in transitory brain states. If this variance is caused by changes in the globally dynamic interplay between distinct functional domains, this can be further investigated in the future.

\section*{Discussion}
\label{sec:Diss}
Search studies in this review demonstrate that the widespread changes in connectivity can observed in post-stroke recovery. In a static brain functional network, decreased interhemispheric FC appears to be a common feature of resting-state network reorganization in stroke, accompanied by reduced network efficiency and modularity. Increased FC can be also observed, and a positive correlation exists between the increased FC of bilateral cerebral hemispheres and the degree of post-stroke functional recovery. On the other hand, the DFC analysis reveals that the FC temporal variability has a decreased trend, and abnormal connectivity states exist in post-stroke patients with motor impairment. Despite DFC analysis is not as mature as SFC in poster-stroke investigation, DFC methods reveal the dynamic nature of brain, bringing an all-new perspective to investigate the recovery mechanism. 

In the following, we continue to discuss two interesting aspects pertinent to SFC and DFC analysis following stroke: 1. if the static or dynamic functional connectivity can sever as a post-stroke recovery biomarker; 2. the methodological considerations relevant to functional connectivity analysis in stroke research.

\subsection*{Static or dynamic functional connectivity as a recovery biomarker following stroke}

Accurate prediction of motor function outcomes and treatment responses after stroke can benefit clinical and research settings, promoting effective rehabilitation care delivery and the stratification of subjects in clinical trials. As a heterogeneous disease, stroke is characterized by the varying lesion size and location. Therefore, the demographic and clinical variables, such as age, sex, lesion volumes, etc., are naturally considered potential factors contributing to the post-stroke recovery prediction~\cite{RN206, RN208}. Recently, there has been increasing interest in the role of functional connectivity measurements acquired from neuroimaging in predicting recovery performance. Hence, in this section, we discuss if the SFC and DFC measurements can serve as a motor recovery biomarker following a stroke from the perspective of FC application.

An appropriate first step to investigating the role of FC measures in motor recovery after stroke is to examine the strength of the association between connectivity and motor behaviour in different stroke populations. A correlation coefficient of 0.75 or greater usually indicates a strong correlation. Results from cross-sectional studies showed a moderate to a strong association between measures of static functional connectivity and motor status after stroke(r =0.58-0.76)~\cite{RN200, RN201, RN199}. The number of stroke patients studied in these studies ranged from 8 to 55. The motion Assessment was measured using the Upper Extremity Fugl-Meyer assessment (UL-FMA) scores, Motricity Index, and Chedoke-McMaster Stroke Assessment.

Regarding DFC analysis, the dynamic FC measure -- temporal variability demonstrates a significant correlation with the UL-FMA scores at the chronic stage after stroke, showing the same effect as static FC analysis~\cite{RN117}. Another DFC study found a negative correlation between the difference in DFC measures in the motor execution network and FMA scores~\cite{RN49}, while it did not pass through FDR correction. In addition, in a study with 31 acute ischemic stroke patients~\cite{RN119}, dynamic functional connectivity pattern shows significant differences between stroke groups varying in motor status: patients with severely impaired mobility are more likely to have a regionally dense connected, highly segregated pattern; patients with mild motor impairment take more time to weakly connect the state with reduced segregation.

The results from the longitudinal studies also appear to underpin FC as a potential biomarker for post-stroke motor recovery. For example, in the longitudinal study of static FC in stroke patients, initial baseline measures of functional connectivity were strongly associated not only with longitudinal temporal assessment scores of motion status~\cite{RN44, RN194}, but also strongly correlated with changes in motor function recovery over time (motor function improvement, r = 0.32 - 0.79)~\cite{RN198, RN196, RN97, RN55}. Dynamic analysis of FC at the longitudinal level further demonstrated the potential of FC alteration as a biological marker of rehabilitation (between bilateral intraparietal lobule and left angular gyrus, r = -0.68)~\cite{RN63}.

With support from the literature findings, FC substantiates that it can serve as a reliable biomarker for post-stroke motor recovery. However, it is advised to keep caution to assess the causality relationship between functional connectivity and stroke recovery— the sample size and statistic capability bring more challenges to FC as a crucial factor in post-stroke recovery. On the one hand, sample sizes collected in clinical analyses typically range from 10 to 20 participants, which is particularly to sensitive to outliers~\cite{RN203} and vulnerable to the effective size inflation~\cite{RN204}, may not accurately represent the entire group. On the other hand, the operation of co-variables, such as age, sex, lesion size/location, baseline motor status/measure, etc., is various across studies, intensifying inconsistent FC recognition in stroke recovery. Nonetheless, we should not be too pessimistic. Despite these criticisms being fair, it does not change the strong or the strong versus weak  FC effects observed in the recovery process.~\cite{RN205}. In the future, longitudinal studies with more samples or the clinical benefits of the function outcome prediction may underpin the role of FC in post-stroke recovery.

\subsection*{Methodological considerations relevant to functional connectivity analysis in stroke research}

Generally, two experimental designs are involved in stroke research with the FC analysis: cross-sectional and longitudinal. These two study designs allow investigation of different post-stroke effects on FC. Typically, a cross-sectional study examines the FC change between stroke-health control (between-person effects); a longitudinal study can peek at the FC change in stroke patients over time (within-person effects)~\cite{RN207}. Due to intensive time and resource payment, the majority of studies rely on cross-sectional designs. However, one of the defects of the cross-sectional study is the brain lesion-induced FC difference will be diluted by the cohort effects. For instance, since the life backgrounds vary in participants, FC differences between stroke patients and healthy will reflect not only the lesion-induced neural circuits reorganization, but also differences in the environment participants live. To track the within-person FC changes in post-stroke recovery and to investigate the causality of FC and neuroplasticity, a longitudinal study is a potent method that can demonstrate the cross-sectional findings from the time dimension~\cite{RN91, RN44, RN55}. In the cross-sectional studies, static and dynamic FC analyses are present, while the static accounts for the majority in longitudinal studies. Recently, dynamic connectivity analysis has been developed in multiple fields~\cite{RN121, RN101, RN114, RN106, RN207}. Since post-stroke recovery is time-dependent, investigating how the neural networks interact dynamically and to what extent dynamic connectivity pattern supports motor function recovery and change deserves further study.

Section II introduces the general pre-processing methods. Although pre-processing steps involve choices between analysis approaches(static vs dynamic), they only vary mildly across FC analysis investigations for post-stroke research. And as far as to our knowledge, the systemic study of the effects of pre-processing choices has been not proposed. That is probably because the sample size of stroke patients is not huge, which cannot support examining the pre-processing steps on the lesion-induced difference in functional connectivity. 

In terms of the brain parcellation methods they use, SFC and DFC studies have apparent preferences. Post-stroke studies with SFC analysis commonly use the atlas-based method to parcellate the whole brain. This brain map generated from massive brain investigation can provide detailed brain information allowing researchers to focus on the interested network or regions. Besides, one of the biggest benefits of using the atlas-based method is that it provides a way to compare and get fairly consistent findings across studies; for example, the lesions interrupt the remote network connection, and the interhemispheric FC decrease, etc. By contrast,  studies with DFC tend to utilize the ICA since the DFC is more sensitive to noise due to increased temporal resolution. Hence, the results of DFC's approach may vary depending on the patient cohorts because the ICA is a data-driven approach which may isolate components belonging to different networks. Moreover, compared with the altas-based method, ICA cannot examine changes in interhemispheric connections. Hence, how connections between hemispheres interact dynamically at the millisecond level may require combining the SFC and DFC results or new intervention approaches.

\section*{Conclusion}
\label{sec:Con}
fMRI based post-stroke functional connectivity analysis for post-stroke motor dysfunction patients has two branches: the static functional connectivity(SFC) and dynamic functional connectivity analysis (DFC). While SFC assumes that the FC/FNC is stationary during the fMRI scan, DFC maintains that the FC/FNC fluctuates even for a short period of time and has a specific coupling pattern. A great many of SFC and DFC analysis methods have been developed and successfully applied to investigate the alteration of functional interaction or communication behind post-stroke motor function deficit and recovery. In this context, this review summarized the current advance of SFC and DFC approaches and latest findings for their application on post-stroke motor function research. The studies included in this review demonstrate that SFC is the predominant post-stroke functional connectivity analysis method in last five fears. The results from SFC show that there is a reduction in FC between motor regions after a stroke, that a rise in FC is highly associated with functional recovery. Meanwhile, DFC is developing rapidly. the application of DFC methods in post-stroke motor function impairment and recovery has shown its potential. With more DFC methods created and utilized to investigate the abnormal motor FC/FNC dynamics, the DFC is expected to far reaching effects in term of neural reorganization underlying stroke recovery and underpin understanding of the recovery mechanism.

Besides, based on the consideration of the previous studies, recommendations from this review of future studies are:
\begin{enumerate}
\item Both SFC and DFC methodology needs to be validated on large cohort to improve reliability and robustness of their statistic results.

\item As a potent method to examine the stroke lesion-effect to FC/FNC dynamics alteration within-person, DFC method based longitudinal post-stroke investigation should be greatly encouraged in the future.

\item The quantitative relationship between FC/FCN alteration and motor function improvement should thoroughly investigated, particular for dynamic FC/FNC alteration.

\item The post-stroke motor function research should not limit to brain motor function areas. The interplay effect between motor network and other function network like the cognitive network should be considered.
\end{enumerate}




\begin{backmatter}

\section*{Acknowledgements}
Text for this section\ldots

\section*{Funding}
Text for this section\ldots

\section*{Abbreviations}

\section*{Availability of data and materials}
Text for this section\ldots


\section*{Competing interests}
The authors declare that they have no competing interests.

\section*{Consent for publication}
Text for this section\ldots

\section*{Authors' contributions}
Text for this section \ldots

\section*{Authors' information}
Text for this section\ldots


\bibliographystyle{mathphys} 
\bibliography{review}      

\end{backmatter}
\end{document}